\documentclass[%
aip,
pop,
amsmath,
amssymb,
preprint,
]{revtex4-1}

\usepackage{graphicx}

\usepackage[utf8]{inputenc}
\usepackage[T1]{fontenc}
\usepackage{mathptmx}

\def\beq{\begin{equation}}
\def\eeq{\end{equation}}
\def\dsp{\displaystyle}

\newcommand{\fig}[1]{Fig.~\ref{#1}}
\newcommand{\Fig}[1]{Figure~\ref{#1}}
\newcommand{\Eq}[1]{Eq.~\eqref{#1}}

\begin{document}
\title{Simulated Refraction-Enhanced X-Ray Radiography of Laser-Driven Shocks}
\author{Arnab Kar}
\email{akar@lle.rochester.edu}
\author{T.~R. Boehly}
\author{P.~B. Radha}
\author{D.~H. Edgell}
\author{S.~X. Hu}
\author{P.~M. Nilson}
\author{A. Shvydky}
\author{W. Theobald}
\author{D. Cao}
\author{K.~S. Anderson}
\author{V.~N. Goncharov}
\author{S.~P. Regan}
\affiliation{Laboratory for Laser Energetics, University of Rochester, Rochester, New York 14623, USA}

\begin{abstract}
Refraction-enhanced x-ray radiography (REXR) is used to infer shock-wave positions of more than one shock wave, launched by a multiple-picket pulse in a planar plastic foil.
This includes locating shock waves before the shocks merge, during the early time and the main drive of the laser pulse that is not possible with the velocity interferometer system for any reflector.
Simulations presented in this paper of REXR show that it is necessary to incorporate refraction and attenuation of x rays along with the appropriate opacity and refractive-index tables to interpret experimental images.
Simulated REXR shows good agreement with an experiment done on the OMEGA laser facility to image a shock wave.
REXR can be applied to design multiple-picket pulses with a better understanding of the shock locations. 
This will be beneficial to obtain the required adiabats for inertial confinement fusion implosions.
\end{abstract}
\maketitle

\section{Introduction}

In direct-drive inertial confinement fusion (ICF) implosions,~\cite{Nuckolls_ICF,Craxton_Review} a spherical target (plastic shell) containing the fusion fuel [deuterium (D) and tritium (T)] is irradiated directly with nominally identical laser beams with the goal of achieving sufficient compression needed for ignition.
Ignition refers to a high gain when the fusion energy from the target is greater than the incident laser energy after the onset on a thermonuclear burn wave.~\cite{Christopherson}
To achieve ignition, the minimum laser energy $(E_{\text{L}})$ and areal density $(\rho R)$ required scale with the adiabat~\cite{Herrmann,Kemp,Betti_adiabat_scaling} or isentrope parameter $(\alpha)$ as $E_{\text{L}} \propto \alpha^{1.88}$ and $\rho R \propto \alpha^{-0.55}$.
The adiabat is a measure of compressibility, defined as the ratio of the electron pressure to the Fermi pressure of the degenerate electron gas at absolute zero temperature.~\cite{Atzeni} 
In the early stage of the implosion, laser light is absorbed by the target and plasma formation occurs through laser--matter interaction.
Typically, the laser pulse comprises a sequence of one to three low-intensity pulses known as ``pickets."
These pickets trigger a series of pressure pulses in the plasma that launch a sequence of shock waves to compress the target.
This sequence, characterized by the individual shock speeds and temporal spacing, creates the desired adiabat profile in the shell necessary for the expected compression.
Adiabat shaping is a technique used in ICF implosions to increase shell compressibility and target yield by increasing the entropy on the outer portion of the shell and preserving a low entropy inside the shell.~\cite{Gardner_adiabat_shaping,Goncharov_adiabat_shaping,Ken_adiabat}
Knowing the position of the shock waves launched over time is crucial to designing the desired adiabat. 

Adiabat shaping in implosions is achieved through hydrodynamic code validation based on shock-timing results from the velocity interferometer system for any reflector (VISAR).~\cite{Barker,Boehly_VISAR}
VISAR provides the speed of the first shock along the axis of propagation and the time at which the other shock waves catch up with the first shock. 
Discrepancies still exist between simulations and experiment for shock velocities and shock-merger time measurements at low adiabats~\cite{Duc} $\alpha \sim$ 1.
This diagnostic does not provide the longitudinal spatial position of the shock waves over time.
Moreover, VISAR does not provide any information about the shock wave early in time because of a time lag associated with the critical surface formation for the diagnostic to work.
During the main drive, the high intensity of the laser leads to x-ray photoionization of the target ahead of the shock front.
This blanks out the VISAR signal, preventing it from determining the shock wave\rq{}s location.~\cite{Boehly_VISAR_blank,Theobald}

X-ray radiography can be a complementary diagnostic in ICF implosions to obtain shock positions by imaging shock waves.~\cite{Hu_Compressibility_Plastic,Montgomery2004,Koch_ICF_refraction,Antonelli}
In this paper, simulated refraction-enhanced x-ray radiography (REXR) has been developed to locate multiple shock waves launched from a multiple-picket pulse.
Phase-contrast imaging, which is widely used in biological and medical imaging,~\cite{Davis,Momose,Mayo,Pfeiffer_Nature} is called refraction-enhanced radiography~\cite{Koch_ICF_refraction,Koch_REI} for Fresnel numbers ($F$) significantly larger than 1.
Here, the Fresnel number is defined as $F=L^2/d\lambda,$ where $L$ is the spatial size of the object and $d$ denotes the distance of propagation of the x rays of wavelength $\lambda$.
For $F \gg 1$, geometrical optics is a good approximation for wave optics that are applicable for imaging typical ICF implosions with soft x rays.
The benefit of this technique over existing x-ray postprocessors when viewing ICF implosions is that it includes x-ray refraction.
For example, \textit{SPECT3D}~\cite{Spect3D,Epstein} accounts only for the attenuation of x rays.
Other potential applications of this method are the study of hydrodynamic mixing at the ablator/fuel interface and the study of mixing of dopants in materials~\cite{Koch_ICF_refraction} along with the effect of the stalk on the spherical targets during implosions.~\cite{Igor}
More recently, it has been shown that inflight density profiles can be inferred for ICF implosions using REXR.~\cite{Dewald}

Through an experimental design, we show that REXR can view shock waves both early in time and during the main drive of the laser pulse, unlike the VISAR diagnostic.
This is possible since REXR does not rely on the formation of the critical surface in early time.
During the main drive, the x-ray photoionization of the target ahead of the shock front does not affect REXR since it relies on the density gradient of the shock front, which is not opaque to x rays.
Additionally, REXR is used to interpret the image of a shock wave for an experiment that was done on OMEGA.
In this context, the importance of the appropriate choice of the opacity and refractive-index tables is discussed in Appendix A.

This paper has been organized as follows: In Sec.~\ref{sec:overview}, an overview of the REXR is presented along with a discussion of the experimental design parameters. 
In Sec.~\ref{simradio}, the results from an OMEGA experiment and simulated REXR are analyzed and compared.
In the experiment, a shock wave was launched inside a planar plastic foil by irradiating it with a laser.
In Sec.~\ref{sec:propose}, an experimental design to image multiple shock waves launched by a picket pulse with a main drive pulse is presented.
Finally, in Sec.~\ref{sec:final}, the capabilities of this diagnostic are summarized.

\section{Overview}\label{sec:overview}

\begin{figure}[!htb]
\centering
\includegraphics[width=0.75\textwidth]{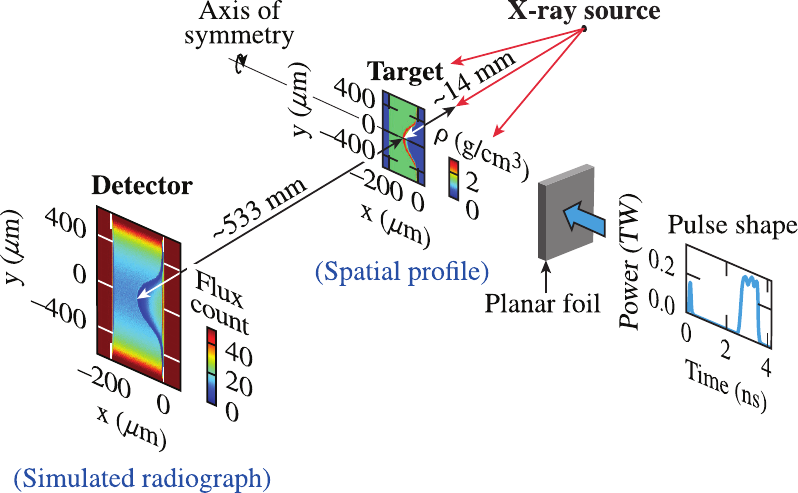}
\caption{A schematic showing a planar foil being irradiated with a laser pulse that generates a shock wave into the target. This process is simulated with the hydrodynamic code \textit{DRACO}, which gives a spatial density profile with the shock features. Point-projected radiography of this profile with x rays of specific energy generates the radiograph on the detector.}
\label{overview}
\end{figure}

Here we present an overview of REXR, which generates images of shock waves.
\Fig{overview} shows a schematic of the different components of this method. 
To generate a shock wave, a laser beam characterized by a pulse shape is incident on a planar foil (ablator) at room temperature, generating shock waves inside the foil. 
This process is simulated with the hydrodynamic code \textit{DRACO}.~\cite{Draco}
Next, x rays sidelight the target at a specific energy and a ray-tracing code tracks the trajectory of the rays as they travel through the density profile obtained from \textit{DRACO} to generate simulated radiographs.
The x-ray energy is chosen such that sufficient fluence to resolve the spatial features is possible. 
The Henke tables provide the refractive indices and opacities corresponding to the density profiles at this x-ray energy.
The simulated radiograph shows the x-ray flux (analogous to the photon count of a detector) by tracking the position of the x rays to account for refraction and their intensity to estimate attenuation.
The details of the ray-tracing code along with a model system to illustrate this idea have been included in Appendix B.
The distances labeled in the schematic determine the magnification of the imaging system.

The experimental design parameters to image a shock wave include the target, the pulse shape that generates the shock waves, and the spatial resolution of the imaging system. 
The relevant ablator materials used in ICF targets include CH and beryllium, which have sufficient x-ray transparency in their compressed states.
The laser drive instrumental in launching (one or more) shock waves determines the degree of compression of the target; then these shock waves are diagnosed with an x-ray framing camera.
The effective spatial resolution of the system is given by the pinhole size.

\section{Comparison of simulated REXR and OMEGA experiments}\label{simradio}

\subsection{Experimental setup}

\begin{figure}[!htb]
\centering
\includegraphics[scale=1.2]{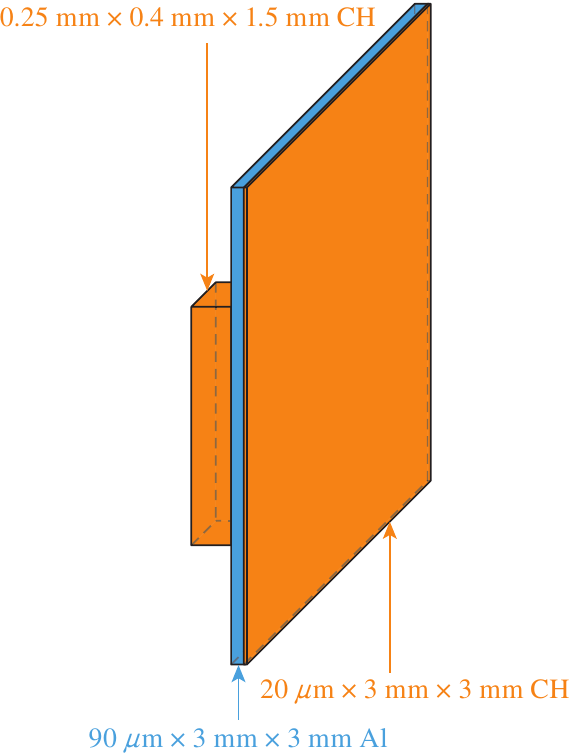}
\caption{The target consisted of a 250-$\mu$m-wide planar plastic foil shielded by an aluminum foil and a thin plastic foil in front of it. The laser beam on the target was incident on the right, driving the shock wave from the thin plastic foil into the aluminum pusher, followed by the thicker plastic foil.}
\label{target}
\end{figure}

\begin{figure}[!htb]
\centering
\includegraphics{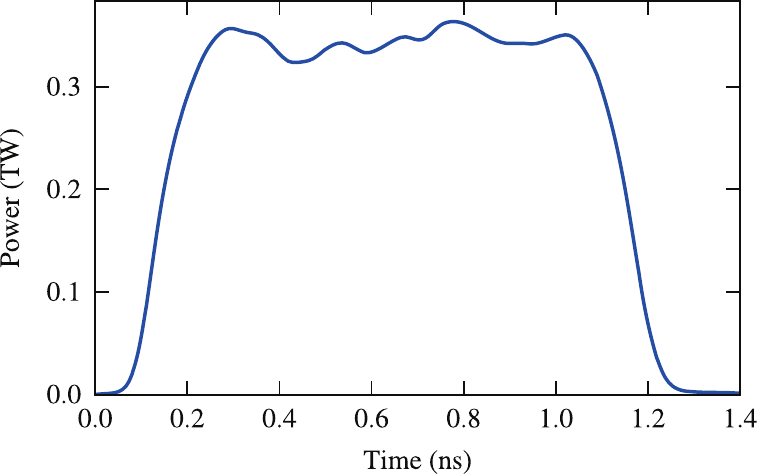}
\caption{A $\sim$1-ns duration square pulse with $\sim$350 J of energy was used for shot 38808.}
\label{laser_drive}
\end{figure}

A point-projection radiography system was used to image a shock wave in a planar plastic foil on OMEGA (shot number 38808).
\Fig{target} shows the target used in the experiment, in which CH was used as the ablator.
To infer the degree of transparency of the x-rays, aluminum was used as a reference material in the target.~\cite{Hicks}
Aluminum also serves as a good shield for the coronal x rays that are generated from the laser drive to enhance the image contrast in the shocked region.
Elements with a low atomic number such as the plastic are used as an overcoat in front of the aluminum foil to provide a higher exhaust velocity and reduce x-ray generation in the corona.~\cite{Ozaki}
A single laser with a $\sim$512-$\mu$m spot size was incident normally on the target. 
The laser drive (shown in \fig{laser_drive}) was comprised of a square pulse with $\sim$350 J of energy that generated the shock wave in the foil.
For the x-ray radiography, x-rays of 5.2-keV energy corresponding to the He-$\alpha$ emissions of Vanadium were projected from a 10-$\mu$m pinhole to image the shock wave onto an x-ray framing camera.
A single strip x-ray framing camera with an integration time of $\sim$300 ps and pixel size of $18\,\mu$m was used. 
The framing camera started to acquire the image at $8.63\pm0.1$ ns after the start of the laser pulse that generated the shock.
The target was placed in the middle, 14 mm away from the pinhole and 533 mm away from the x-ray framing camera.

\subsection{Simulated REXR}

\begin{figure}[!htb]
\centering
\includegraphics{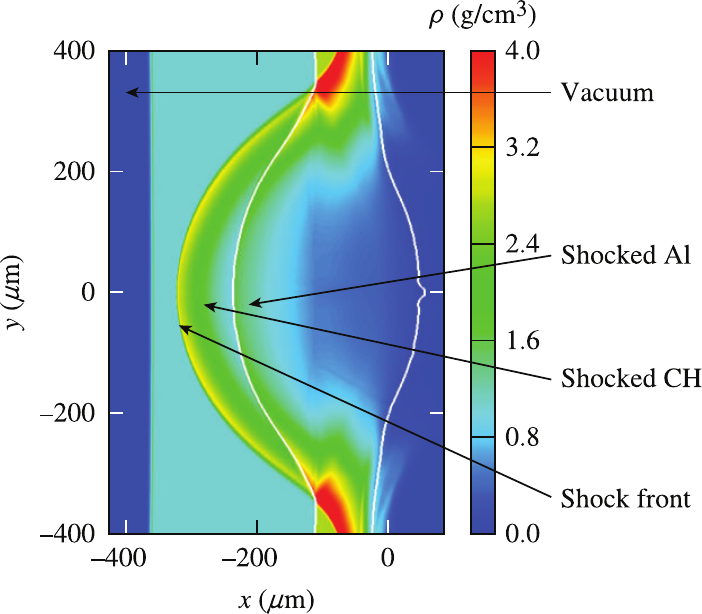}
\caption{Density profile at 8.62 ns obtained from a \textit{DRACO} simulation modeling an experiment on OMEGA. Along the center of the beam axis, the highest-density point at --320 $\mu$m marks the shock front in plastic that was launched from the point of the incident laser energy at 0 $\mu$m on the right. The solid white line corresponds to the plastic/aluminum interface.}
\label{draco_38808}
\end{figure}

\textit{DRACO} simulations for this experimental setup were performed.
A Eulerian version of the code modeled the laser energy deposition on the ablator through ray-tracing and included processes such as heat conduction and radiation transport.~\cite{Draco}
\textit{DRACO} is an axially symmetric code by nature, and the shock waves propagated along the center of the beam axis in these simulations.
\Fig{draco_38808} shows the density profile that was obtained from the simulation at 8.62 ns when the target was no longer being irradiated by the laser beam.
The interface between the vacuum and rear end of the plastic is $360\,\mu$m from the initial surface upon which the laser was incident.
Near the surface of the laser deposition, the figure shows plastic and aluminum of low density as they were ablated. 
This was followed by a gradual rise in the density of the aluminum caused by the transition of the shock wave.
The density gradually increased in space as it approached the shock front in the plastic region.
The highest-density point in plastic corresponding to the shock front occurs around $-320\,\mu$m.
The shock profile in plastic shows a prominent bowing effect because of the smaller laser spot size relative to the target.
The shock wave was not being supported by then as the laser drive was turned off.

\begin{figure}[!htb]
\centering
\includegraphics{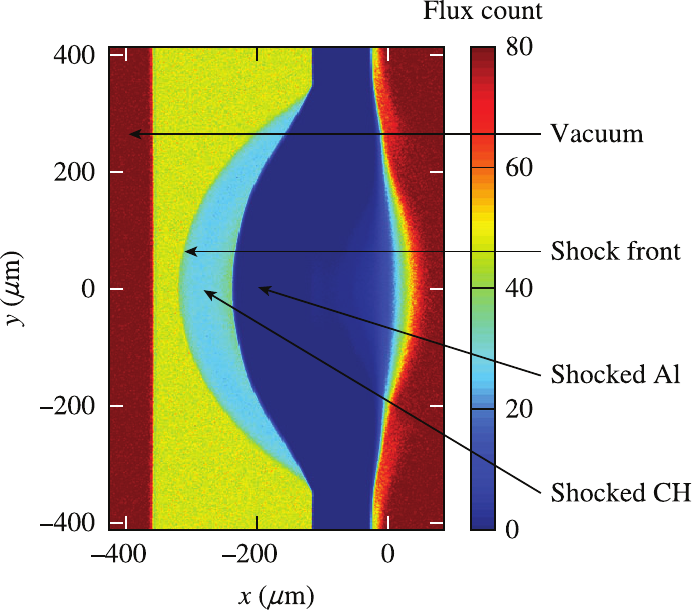}
\caption{Simulated radiograph for an OMEGA experiment shows the shock profile in a plastic ablator. The x-ray flux is representative of the degree of transmission of the x rays through the different areas: vacuum (in red), unshocked plastic (in yellow), shocked plastic (in cyan), and shocked aluminum (in blue). The color contrast in the figure shows the shock profile distinctly.}
\label{radio_38808}
\end{figure}

Subsequently, x rays of 5.2-keV energy were used to generate the simulated radiograph in \fig{radio_38808} using refraction-enhanced radiography.
The image at the detector is a magnified version of the figure shown here in the target plane, where the magnification is determined by the ratio of the distance between the detector and pinhole to the distance between the target and pinhole.
For simulated radiography, the Henke tables determine the refractive indices and the opacity of the medium corresponding to the density profiles obtained from the hydrodynamic code.
The reason behind the particular choice of this refractive-index/opacity table is provided in Appendix A.
The highest x-ray flux in the vacuum region of the image shows that the x rays do not get attenuated here. 
The x-ray flux represents the degree of transmission of the x rays in these regions that produces the contrast in the radiograph so that the shock profile is visible.
The transmission that is dependent on the density conditions and the material opacities gradually increases as the shock wave transits. 
The bowing effect of the shock wave is also prominently visible in the radiograph.

\begin{figure}[!htb]
\centering
\includegraphics{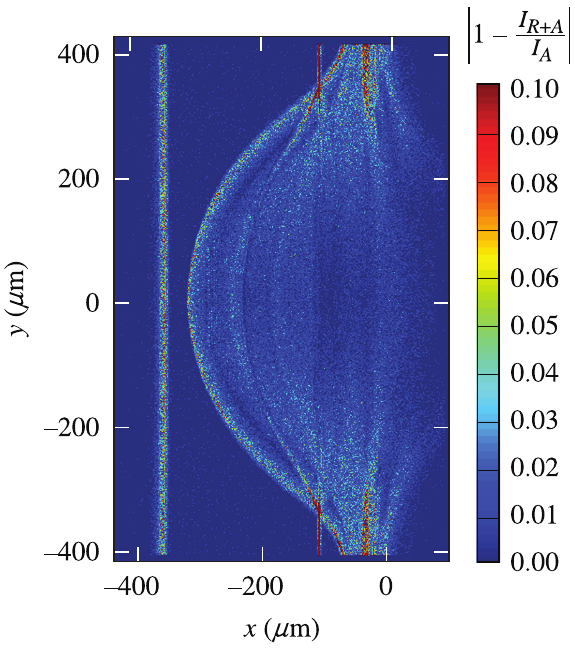}
\caption{The effect of refraction on the x-ray flux is shown here by taking the normalized difference in two radiographs: one that includes refraction of x rays ($I_{R+A}$) and one that does not ($I_A$). The figure shows that the refraction of x rays leads to changes in the x-ray flux near the shock front and the vacuum/plastic interface.}
\label{refrac_effect}
\end{figure}

Due to the bowing effect of the shock front, it is important to account for the refraction of the x rays.
Moreover, the sharp density gradient across the shock front leads to the deflection of the x rays because of the steep gradient in the refractive indices.
To see the contribution of the refractive effects specifically, the density profile from \textit{DRACO} was radiographed by two methods:
In the first method, both refraction and attenuation ($I_{R+A}$) of the x rays were included, as seen in \fig{radio_38808};
in the second method, only attenuation ($I_A$) of the x rays was included.
\Fig{refrac_effect} shows the difference in the normalized flux $\left(|1-I_{R+A}/I_A|\right)$ between the two radiographs.
The difference in the x-ray flux in the bow-shaped shocked region and the material interfaces is evident, suggesting the deflection of the x rays.
In the aluminum region, the x rays have very low intensities and the relative differences are not noticeable.
This deflection of the x rays across the vacuum/plastic interface was seen in the experiment, as discussed later.

\subsection{OMEGA experiment}

\begin{figure}[!htb]
\centering
\includegraphics{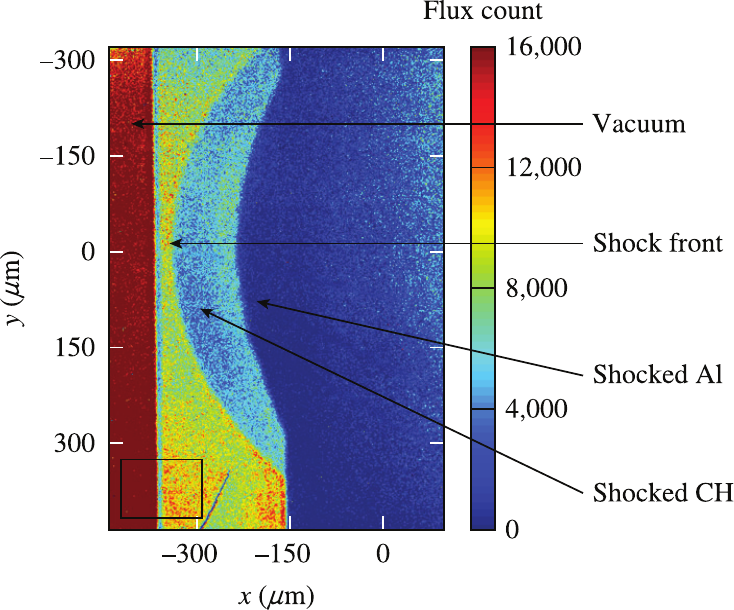}
\caption{Image obtained from an x-ray framing camera on OMEGA. The image was captured around 8.63 ns after the target was irradiated with a square pulse that launched the shock wave. The image shows the bowing effect of the shock wave in plastic with the main features labeled. The region marked with a black box shows the vacuum/unshocked plastic interface. Lineouts from this region shown in \fig{lineout_38808} are used to infer the opacity of unshocked plastic as a consistency check.}
\label{draco_fig}
\end{figure}

\Fig{draco_fig} shows the experimental image of the shock wave inside a planar plastic foil that was obtained on OMEGA.
The unshocked CH, shocked CH, and aluminum pusher are easily distinguishable in the image because of the difference in the degree of transmission of the x-rays in these regions.
The bowing effect of the shock front is also prominent.
It should be pointed out that the unshocked plastic region in the image extends over $\sim$200 $\mu$m in the direction of the shock-wave propagation; however, it should have been $250\,\mu$m as per the experimental design.
One possible reason why we do not see the full extent of the plastic in the image is an error in target alignment.
It is possible that the x rays were not incident normally on the target and were attenuated by the large aluminum pusher foil in the target. 
This would also account for the sharp jump in the flux count of the shock profile at the plastic/aluminum interface. 
Instead, a smooth bow-shaped shock front was expected at the aluminum/plastic interface as we see in the \textit{DRACO} simulations (see \fig{radio_38808}).
Overall, this experiment shows that dynamic shock waves can be imaged using this technique. 

\begin{figure}[!htb]
\centering
\includegraphics{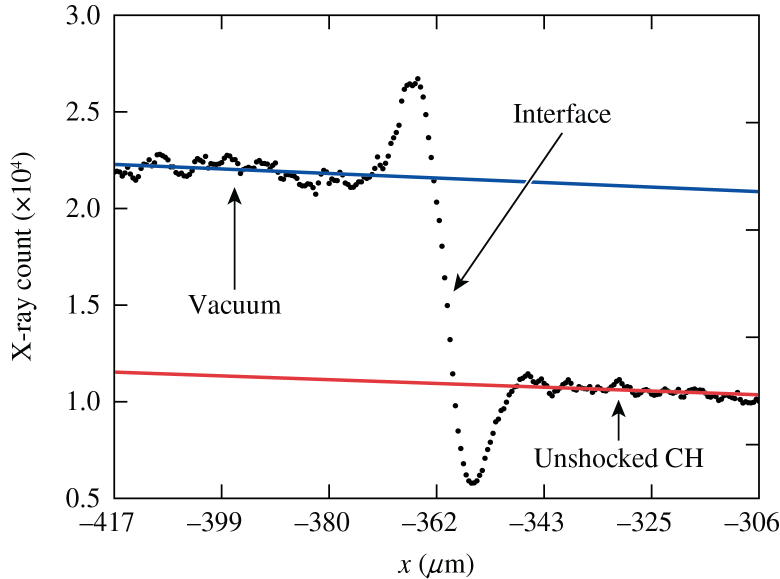}
\caption{Lineouts across the vacuum/plastic interface from \fig{draco_fig} are shown here. The red and blue lines represent the linear fits to the x-ray flux in the vacuum and unshocked CH respectively. The peak and valley in the lineout show the effect of the refraction of the x rays across the CH/vacuum interface.}
\label{lineout_38808}
\end{figure}

For verification purposes, a simple check to infer the opacity of the unshocked plastic region was performed. 
Lineouts from the interface of the vacuum and unshocked CH region marked with a black box in \fig{draco_fig} are shown in \fig{lineout_38808}.
The x-ray flux in the vacuum is high since there is no attenuation.
At the interface, the x-ray flux fluctuates because of refraction of x rays across the plastic/vacuum interface, making the x rays deflect from their trajectory of propagation.  
In the plastic, the x-ray flux decreases because of its higher opacity compared to the vacuum.
In fact, these x-ray flux can be used to quantify the opacity of plastic using the relation
\beq\label{eq:opac}
I=I_0 e^{-\kappa_{\text{CH}} L}, 
\eeq
where $I, I_0$ are the x-ray flux in plastic and vacuum, respectively, $L$ is the distance of propagation of the x rays through the plastic foil, and $\kappa_{\text{CH}}$ denotes the opacity of unshocked plastic. 
In the experiment, $L=400\,\mu$m, i.e., the depth of the plastic foil.
From the intercepts of the linear fits, $I_0=2.23\times10^4$ and $I=1.15\times 10^4$ are obtained.
Using these values in \Eq{eq:opac}, the opacity of the unshocked plastic is inferred to be 16.5 cm$^{-1}$.
This matches well against the opacity of 16.3 cm$^{-1}$ for plastic of 1.05-gm/cm$^3$ density as per the Henke tables at a photon energy of 5.2 keV.

\subsection{Comparison of simulation to experiment}

\begin{figure}[!htb]
\centering
\includegraphics{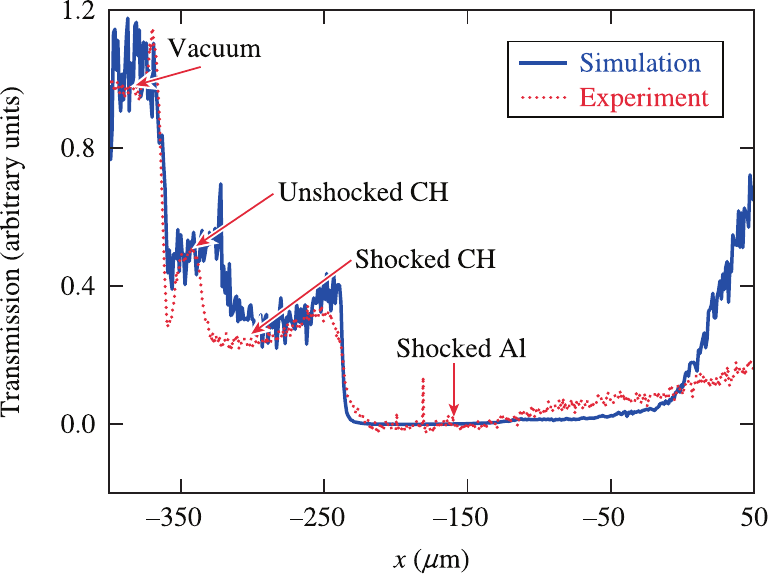}
\caption{The transmission curves along the center of the beam axis were obtained from the simulated radiograph and the experimental image which show good agreement between them for plastic and aluminum. For reference, the transmission in the vacuum region is set to 1 since there is no attenuation.}
\label{lineout}
\end{figure}

To compare the experimental and the simulation results, the transmission curves have been compared in \fig{lineout}.
The lineouts have been taken about the center of the beam axis of Figs.~\ref{radio_38808} and \ref{draco_fig}.
Both the lineouts have been scaled so that the transmission through the vacuum region is $\sim$1. 
We begin by inferring the opacity of the unshocked plastic obtained from simulation.
By using \Eq{eq:opac} with $I=0.51$ and $I_0=0.99$, $\kappa_{\text{CH}}$ = 16.7 cm$^{-1}$.
This inferred opacity matches the experimental data and atomic physics calculations of the Henke tables for x rays of 5.2-keV energy. 
The plot shows that the relative degree of transmission in the unshocked plastic, shocked plastic, and shocked aluminum with respect to the vacuum is in good agreement between the experiment and refraction-enhanced radiography.
The actual position of the shock front (adjacent to unshocked CH on the right in the figure) and the spatial extent of the shocked regions are in reasonable agreement with each other. 
A plausible explanation for the lack of an exact match in the spatial extent of the shocked regions is caused by lateral heat flow from the decaying shock since the image was taken very late in time after the laser pulse had been turned off.
Typically, experimental investigations of laser-driven shocks occur within a few hundred picoseconds at the end of the laser pulse, unlike this specific experiment.
Therefore, any disagreement in the drive (that determines the shock strength) after the pulse was turned off will be amplified over time. 
Near the ablation front, the transmission profiles do not agree well. 
The target alignment error is the likely reason for this mismatch. 
A small tilt in the large aluminum foil, when projected onto the 2-D image, can blank out the signal over a significant region.
However, in the ablation region, the results from the simulated radiograph and the hydrodynamic simulations are consistent with each other (comparing Figs.~\ref{draco_38808} and \ref{radio_38808}).
The discrepancy in the ablation region could also occur from a lower mass ablation rate in simulations compared to the experiment.
Overall, the good agreement between the relative degree of transmission between refraction-enhanced radiograph and the experiment gives us the confidence to rely on our modeling capability.
The target alignment error and the imaging at a late time after the pulse is turned off discussed above affect the spatial extent of the shocked/unshocked regions of a material but do not alter the relative degree of transmission in a material with respect to the vacuum.

It should be mentioned that for refraction-enhanced x-ray radiography, it is not possible to infer a 1-D density profile (lineout) accurately from the transmission profile. 
This is because of the x-ray refraction and the bowing of the shock wave, which is inherently a 3-D phenomenon.
Even if the shock curvature is assumed to be axially symmetric and thereby account for a variable density (therefore, opacity) along the path length, the specific shape of the post-shock density profile at different radii from the beam axis will determine the deflection of the x rays caused by refraction.
In regular absorption radiography that infers a path-integrated density for a straight line ray path, x-ray refraction is neglected. 
However, a recent paper~\cite{Dewald} has shown that inflight density gradients can be inferred for spherical implosions.

\section{Application of REXR to proposed experiments}\label{sec:propose}

In this section, we propose an experimental design to image multiple shock waves with refraction-enhanced radiography. 
The objective is to view and locate the shock position during the main drive pulse when the experimental diagnostics such as VISAR fail to locate the shock position.
Due to the high intensity during the main drive, x-ray photoionization of the target ahead of the shock front leads to absorption of the optical diagnostic of the VISAR and no signal is generated.~\cite{Theobald,Boehly_VISAR_blank}
This analysis will be useful for understanding the dynamics of shock coalescing and designing pulse shapes with multiple-picket pulses to achieve the desired adiabats.~\cite{Goncharov_adiabat_shaping,Ken_adiabat}
Such an analysis will also provide additional information about shock locations before they merge. 
This information will improve the way pulse shapes are currently designed based on VISAR data of shock-merger experiments.
This may also shed some light on why the shock-merger experimental data do not agree well with theoretical estimates for low adiabats.~\cite{Duc}

\subsection{Picket pulse preceding the main target drive pulse} 

\begin{figure}[!htb]
\centering
\includegraphics{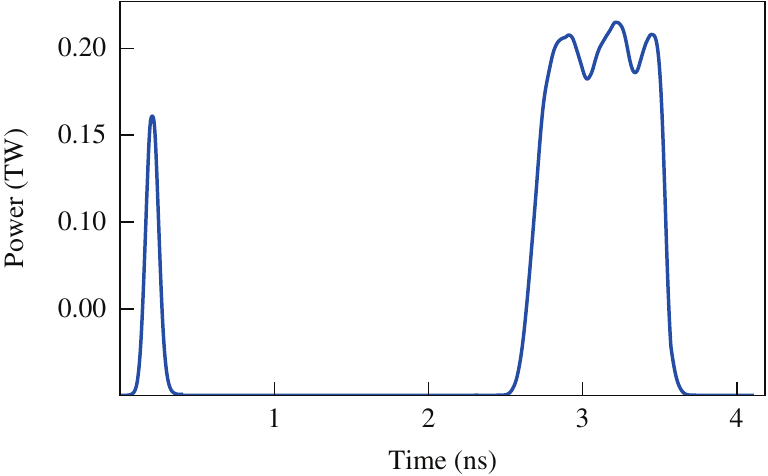}
\caption{A single-picket pulse is shown with a main drive pulse of 190-J energy. A pulse shape of this kind launches two shock waves whose positions are analyzed in this section.}
\label{pick_pulse}
\end{figure}

\begin{figure}[!htb]
\includegraphics{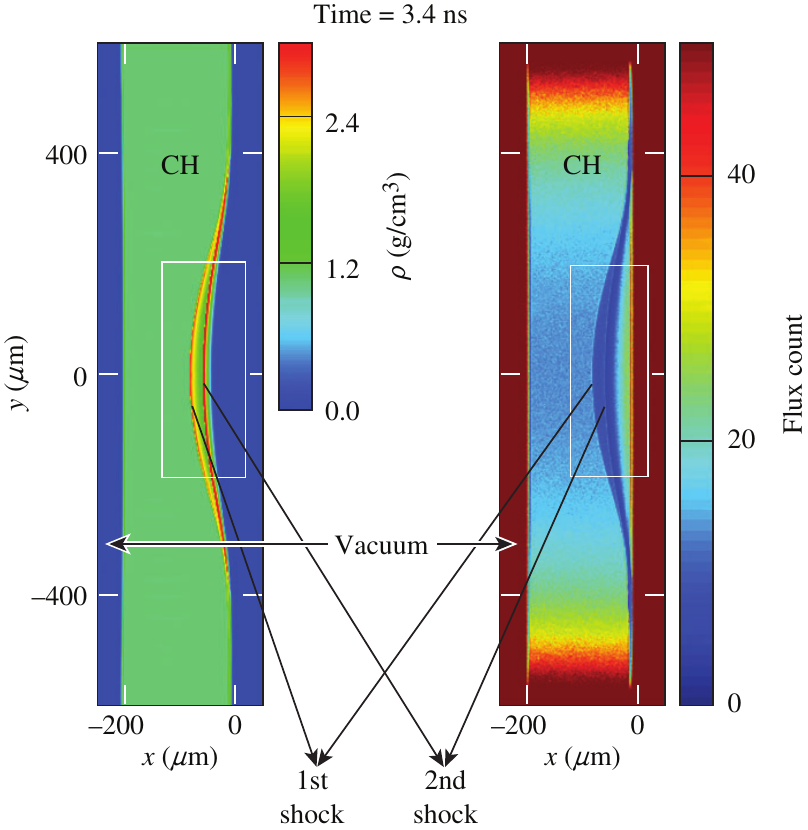}
\caption{The density profile from a \textit{DRACO} simulation (on left) and the corresponding radiograph (on right) are shown. The white boxes mark the region of interest where the two shock waves caused by the picket and the main drive pulse are labeled. To see the spatial features clearly, only this region of interest will be shown in \fig{pick_all} for five time steps.}
\label{pick_full}
\end{figure}

In this design, a picket with a main drive pulse as shown in \fig{pick_pulse} is used to drive shock waves through a square planar plastic foil of 1.2-mm height and 200-$\mu$m width.
This setup is modeled with a \textit{DRACO} simulation that provides the density profiles over time.
One such \textit{DRACO} profile along with the corresponding simulated radiograph is shown in \fig{pick_full}.
The two shock fronts in plastic due to the picket pulse and main drive are distinctly visible. 
The degree of attenuation changes with the density conditions in the plastic, which is captured by the radiograph.

\begin{figure*}[!htb]
\includegraphics{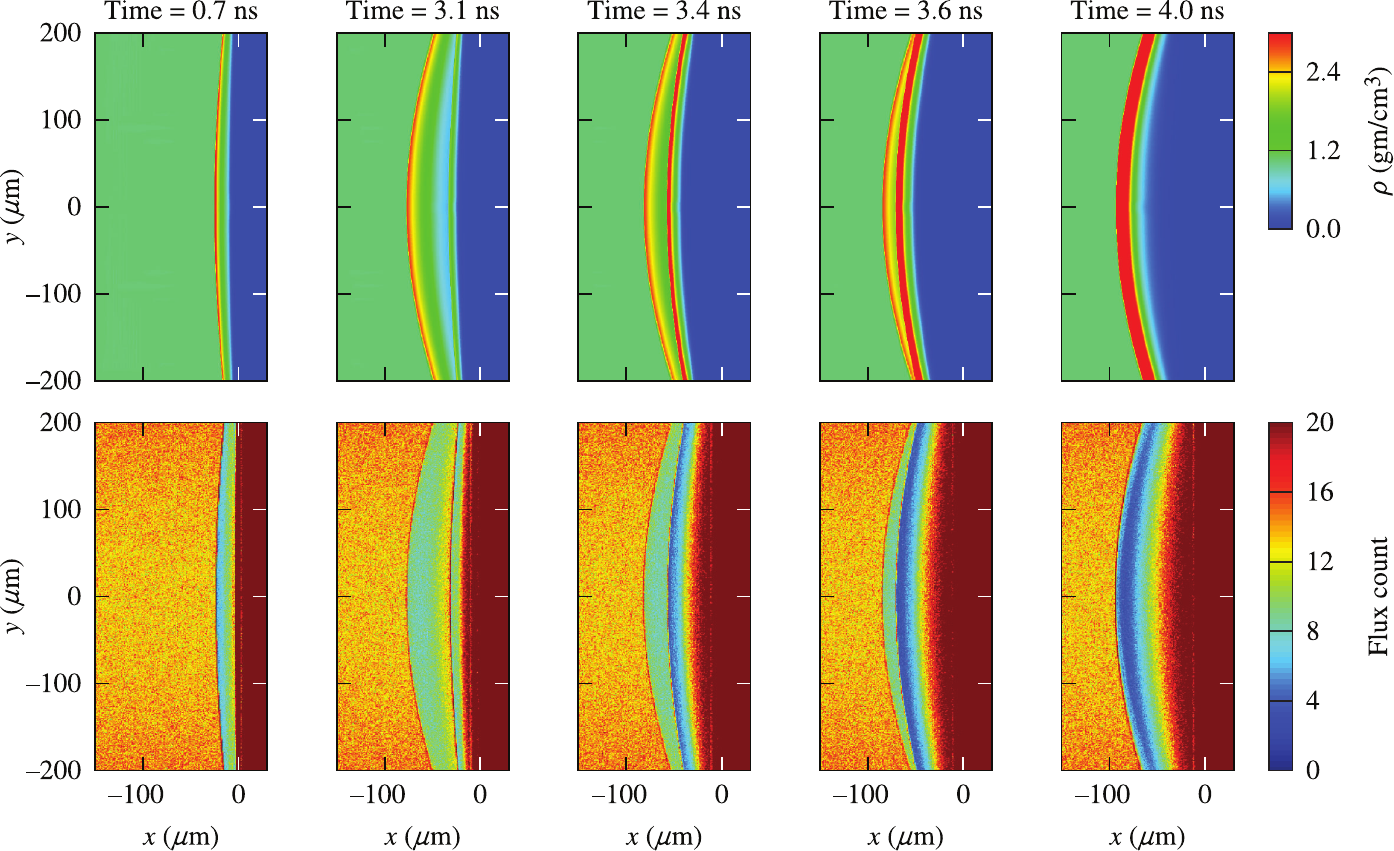}
\caption{The spatial profiles obtained from \textit{DRACO} simulations and the corresponding simulated radiographs are shown here over time. With increasing time (going from left to right), the second shock wave launched during the main drive catches up with the slow-moving first shock wave launched from the single picket. The bowing effect of the first shock becomes more prominent over time compared to the second shock wave, which is launched later.}
\label{pick_all}
\end{figure*}

\Fig{pick_all} shows the evolution of the spatial profiles and the corresponding simulated radiographs over time. 
These profiles highlight the regions of interest, although the radiographs were generated over a larger area.
The figure shows the second shock wave that was triggered late in time catching up with the slow-moving first shock wave. 
This analysis provides the spatial information about the distance between the shock waves before they merge. 
It shows that by the time the second shock wave is detectable around 3.1 ns, half of the energy from the main drive has already irradiated the target.
The shocks merge around 4 ns, which is $\sim$0.4 ns after the main drive has been turned off. 

\begin{figure*}[!hbt]
\includegraphics{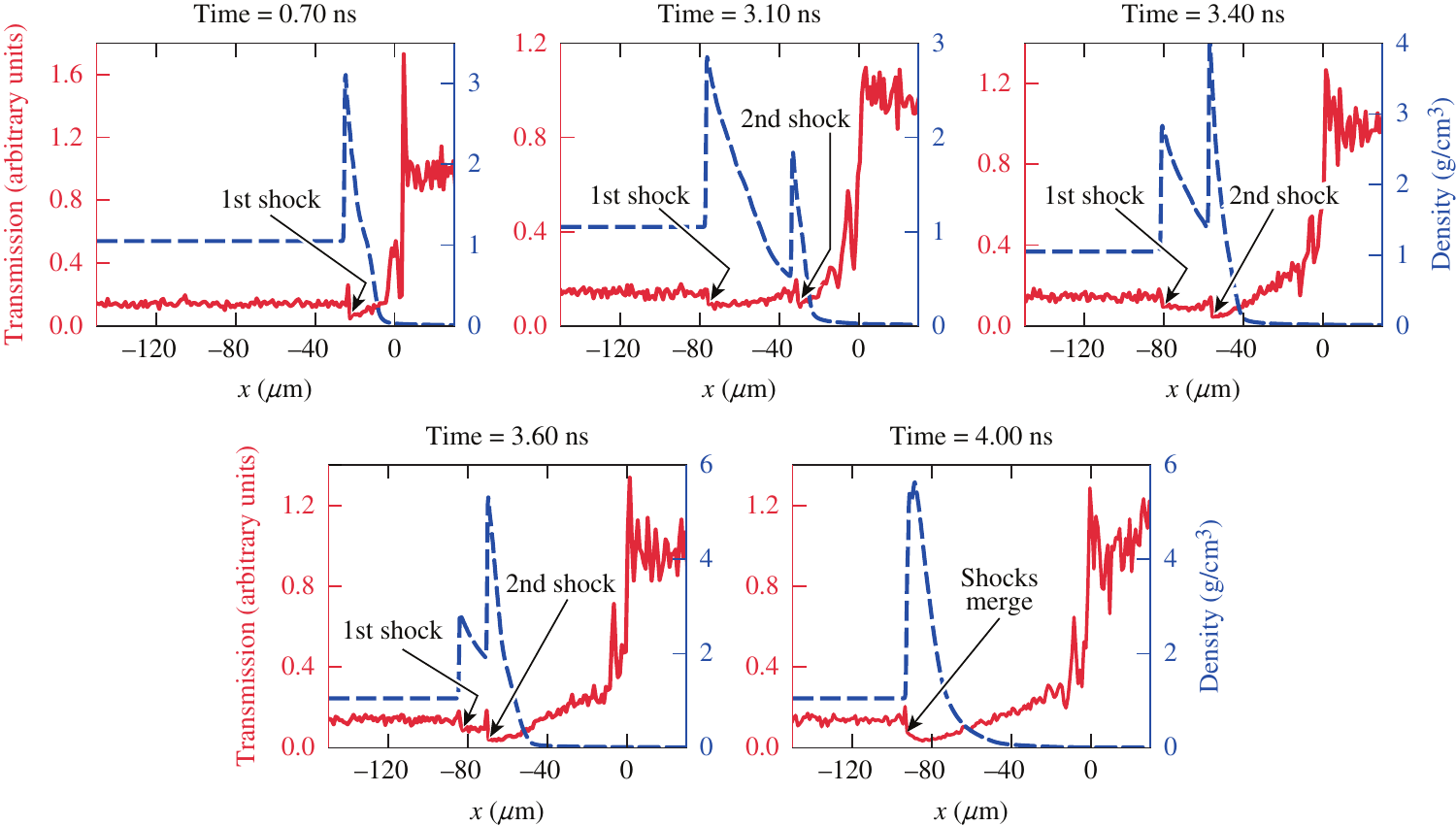}
\caption{The transmission (red) and density profiles (blue) across the center of the beam axis obtained from \fig{pick_all} are shown here. The transmission has been scaled so that the intensity in the vacuum region is 1. The spikes followed by the dip (local minimum) in the transmission curve correspond to the shock fronts as labeled. The density profile also spikes at those points to illustrate this fact.}
\label{pick_line_all}
\end{figure*}

The density profile and degree of transmission of the x rays through these shock profiles about the center of the beam axis are shown in \fig{pick_line_all}.  
As expected, the transmission of the x rays reflects the density conditions in the shock waves and the post-shock density conditions. 
The spike and dip in the transmission correspond to the position of the peak densities of the shock waves.
The spike and dip occur due to the refraction of the x rays across the shock front.
The position and shape of these features are dependent on the density profiles and are not affected by the finite size of the pinhole in the imaging system.
The local minimum (or drop) in the transmission occur due to the lowest transmission of the x rays through the relatively highest-density condition (local maximum) across the shock front.
The position of the shock front can be inferred from these drops by locating the local minimum in the transmission curves. 
The inferred shock positions from the simulated radiographs were found to be in good agreement with shock locations defined by the density profiles.
The fluctuations near the ablation front show that the x-ray trajectories get deflected significantly in this region.
This happens due to the changes in the density profile in the lateral direction at different radii from the beam axis as a result of the ablation.
In the experimental image shown previously, this was absent due to absorption in aluminum.
To summarize, the shock-wave transit during the main drive is noticeable before it catches up with the first shock launched from the picket pulse around 4.0 ns.

\section{Conclusion}\label{sec:final}

We have developed a method to generate images of shock waves that are relevant for ICF implosions through refraction-enhanced x-ray radiography.
This method is capable of inferring shock-wave position, both during the early stage of the laser pulse and the main drive, which is not possible with the current experimental diagnostics such as VISAR.
The point-projection backlighting described in this paper provides a transverse 2-D image of the shock waves from which their positions can be inferred for more than one shock wave before they coalesce. 
The usefulness of the technique was illustrated by generating images of multiple shock waves and determining their locations.
The calculations indicate that it is important to account for the refraction of x rays across density gradients of the shock fronts and material interfaces along with their attenuation and to choose appropriate opacity tables to interpret experimental observation.
Efforts are underway to use REXR as a diagnostic for OMEGA experiments to image shock positions and mergers. 
This post-processor will be used to identify observables for the experiments that will be enhanced due to the refraction in shock profiles.

\begin{acknowledgments}

We would like to thank T.~J.~B. Collins, L. Antonelli and F. Barbato for useful discussions. 
This material is based upon work supported by the Department of Energy National Nuclear Security Administration under Award Number DE-NA0003856, the University of Rochester, and the New York State Energy Research and Development Authority. 

This report was prepared as an account of work sponsored by an agency of the U.S. Government. Neither the U.S. Government nor any agency thereof, nor any of their employees, makes any warranty, express or implied, or assumes any legal liability or responsibility for the accuracy, completeness, or usefulness of any information, apparatus, product, or process disclosed, or represents that its use would not infringe privately owned rights. Reference herein to any specific commercial product, process, or service by trade name, trademark, manufacturer, or otherwise does not necessarily constitute or imply its endorsement, recommendation, or favoring by the U.S. Government or any agency thereof. The views and opinions of authors expressed herein do not necessarily state or reflect those of the U.S. Government or any agency thereof.

\end{acknowledgments}

\appendix 

\section{Sensitivity of REXR to opacities and refractive indices}

REXR is sensitive to the refractive indices and opacities of the materials and the x-ray photon energy.
X-rays traveling through a medium are refracted and attenuated. 
The refractive index of a material $(n)$ has a real and imaginary part.
If we denote $n=(1-\delta)+i \beta$, then Re$(n)=(1-\delta)$ leads to refraction and Im$(n)=\beta$ controls attenuation.
Specifically, the opacity of a material $(\kappa)$ is related to the imaginary part of the refractive index through the relation $\kappa=4\pi\beta/\lambda$, where $\lambda$ is the x-ray wavelength.

\begin{figure}[!htb]
\centering
\includegraphics{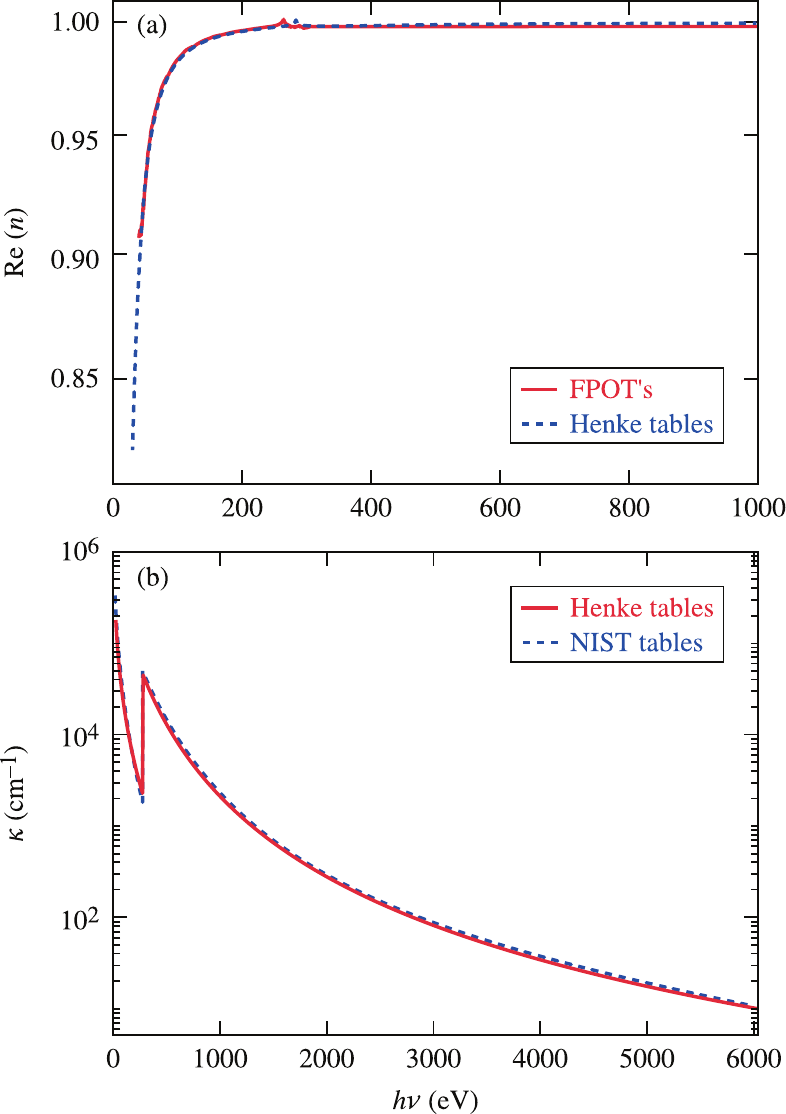}
\caption{
(a) The real part of the refractive index Re$(n)$ obtained from the first-principles opacity tables (FPOT\rq{}s) and the Henke tables is compared for x-ray photon energy between 40 eV and 1 keV. 
(b) The opacity $(\kappa),$ which is related to the imaginary part of the refractive index $(\kappa=4\pi\beta/\lambda),$ is compared between the Henke tables and the NIST tables for x-ray photon energy between 30 eV and 6 keV. In this plot, the NIST data and Henke table data correspond to room temperature values of 1.05-gm/cm$^3$ CH while the FPOT values are at 5000 K.}
\label{comparison}
\end{figure}

To incorporate the material properties accurately, the choice of the refractive indices and the opacity tables play a crucial role.
The first-principles opacity tables~\cite{Suxing_opacity,Hu_Review_PoP} (FPOT\rq{}s) and Henke tables~\cite{Henke} are two databases that can be used [by inferring opacity from Im$(n)$] since they both provide the Re$(n)$ and Im$(n)$ of the refractive indices. 
Some of the standard choices for the opacity tables, such as the astrophysical tables,~\cite{AOT} \textit{PROPACEOS} tables,~\cite{PROPACEOS} and NIST tables,~\cite{NIST} do not provide the real part of the refractive index.
It is worth mentioning that exact quantum mechanical calculations that have been done for x-ray energies of less than 500 eV for the FPOT\rq{}s~\cite{Hu_Drude} are in good agreement with the standard NIST opacities for cold materials.
Beyond that for higher-energy x rays, the values in the FPOT\rq{}s are extrapolated using the Drude model. 
It is noticed that the agreement between the extrapolated values from FPOT\rq{}s and the standard NIST cold opacities differ by a larger amount (not shown here). 
Taking these into consideration, the Henke tables are the only available choice to obtain the density-dependent Re$(n)$ and Im$(n)$ for different materials.
To check the reliability of the Henke tables, the Re$(n)$ is compared to the refractive indices obtained from the FPOT\rq{}s.
The opacity corresponding to Im$(n)$ of the Henke tables is compared to the NIST opacity.
\Fig{comparison} shows good agreement between the Henke tables and the above-mentioned databases. 
Since plastic (CH) of 1.05- gm/cm$^3$ density has a plasma frequency of $\omega_{\text{p}} <$ 30 eV, the refractive index increases beyond that and approaches 1 for x-ray energies $\gg\omega_{\text{p}}$. 
Around photon energies of 280 eV, both the refractive index and the opacity are discontinuous because of the K edge of carbon in CH when the 1$s$-core electrons become accessible, which increases the electrical conductivity.~\cite{Suxing_opacity}
It should be noted that the discontinuity in the Re$(n)$ around photon energy of 280 eV does not coincide exactly between the FPOT\rq{}s and the Henke tables. 
This happens because of the underestimation of the band gap (or correspondingly the K edge) in the density functional theory calculation that generates the FPOT\rq{}s.
Since Re$(n)$ and $\kappa$ are related to electrical conductivity, their values are discontinuous; but overall, the opacity of CH monotonically decreases after the K edge.

Intuitively, material properties such as refractive index and opacity under shocked conditions should be different from room-temperature conditions. 
To check this, we compared the Re$(n)$ from the FPOT\rq{}s and opacity from the astrophysical tables for CH at solid density for different temperatures.
It was observed that temperature does not influence the refractive indices and opacities significantly so its dependence can be neglected. 
Nevertheless, it should be noted that there is a K edge shift in the opacity values with a rise in temperature,~\cite{Suxing_opacity} but the K edge for C in plastic is at 284 eV while the x-ray imaging for shock-wave radiography is performed around 5 keV.

The x-ray photon energy is another parameter that affects the refractive indices. 
Specifically, the real and imaginary parts of the refractive index scale as~\cite{Mayo,Fitzgerald} 
\beq
\delta \propto (h\nu)^{-2}, \quad \beta \propto (h\nu)^{-4},
\eeq
with the x-ray photon energy $h\nu$.
This leads to a competition between the refractive and attenuative effects based on the x-ray energy.
As the x-ray photon energy increases, the refractive effects become less dominant as the refractive indices approach the refractive index of vacuum, i.e., 1 (since $\delta \to 0$).
However, the attenuation is decreased significantly and plastic becomes more transparent to x rays.
To obtain sufficient fluence from the x rays to resolve the spatial features, this energy dependence is an important consideration when determining the optimal x-ray energy in the radiography.

\section{Refraction-Enhanced X-Ray Radiography}

Here, we discuss the details of the refraction-enhanced x-ray radiography (REXR). 
X rays are launched in a point-projection radiography setup at a specific energy, and their trajectories are tracked as they pass through the material profiles obtained from \textit{DRACO}.
The x rays are modeled as rays and their trajectories are determined through the ray equation described below.
The ray-tracing approach or geometrical optics is applicable in our setup and a wave-based treatment is not necessary.
This is justified by the fact that the Fresnel number ($F$) of the setup is significantly larger than 1.
Fresnel number $F$ is defined as $F=L^2/d \lambda$, where $L$ is the characteristic size of the spatial features and $d$ is the propagation distance of the x rays of wavelength $\lambda$.
For $F \gg 1$, phase-contrast imaging is referred to as refraction-enhanced radiography.

In geometrical optics, the path of a ray traveling through a medium of varying refractive index is described by the following differential equation:~\cite{Born}
\beq\label{arc}
\frac{d}{d s}\left[n({\bf r})\frac{d {\bf r}}{d s}\right]=\nabla n({\bf r}),
\eeq
where ${\bf r}$ denotes the position vector of a point on the ray, $d s$ is an element of the arc length on the ray, and $n({\bf r})$ represents spatial dependence of the refractive index of the medium.
Through a change of variables $dt=d s/n$, \Eq{arc} can be written as 
\beq
\frac{d^2 {\bf r}}{d t^2}=n({\bf r}) \nabla n({\bf r}).
\eeq
This second-order equation can be further simplified into two first-order differential equations by defining an optical ray vector ${\bf T}$:
\beq\label{sol}
{\bf T}=\frac{d {\bf r}}{dt}, \quad \frac{d {\bf T}}{d t}=\frac{1}{2}\nabla n^2({\bf r}).
\eeq
This pair of ray-tracing equations is solved through a Runge--Kutta method described in Ref.~\onlinecite{Sharma}.
The refractive index in \Eq{sol} is obtained from the Henke tables based on the material densities.
The method also accounts for attenuation of the x rays by tracking the cumulative decrease in the intensity of the x-ray along its path length through exp$\left[-\dsp{\int \kappa({\bf r}) ds}\right]$ where $\kappa({\bf r})$ is the opacity distribution of the medium. 
Henke tables provide the imaginary part of the refractive index, i.e., $\beta$ and the opacities are determined from $\beta$ by using $\kappa=4 \pi\beta/\lambda$.
The final position of the x rays in the detector plane and their intensities after they pass through the spatial profile are determined from the ray-tracing equations.
The final positions account for the amount of refraction or the extent of deflection of the x rays and their intensity determines the attenuation.

\begin{figure}[!htb]
\centering
\includegraphics{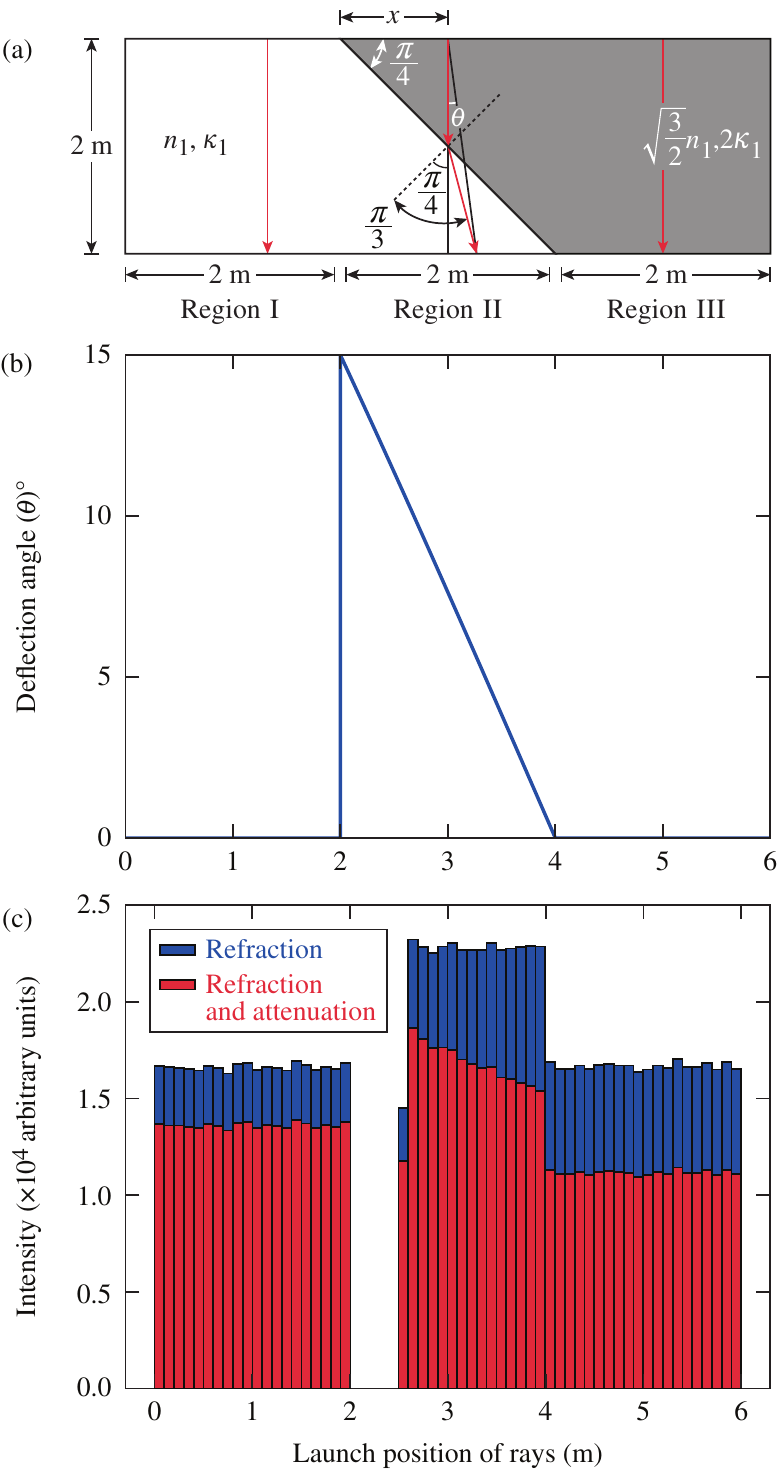}
\caption{
(a) A rectangular slab with two materials having different refractive indices and opacities is shown along with the path of representative rays (in red) in the three regions of the slab. 
(b) The angle of deflection of the rays from their initial direction of propagation is plotted as a function of their launch position.
(c) The intensity of the rays shows the effect of refraction (or refraction and attenuation) by plotting the number of rays with unit intensity \{or intensity $\dsp{\exp\left[-\int \kappa({\bf r})ds\right]}$\} as a function of their launch position.}
\label{test}
\end{figure}

Here, we have outlined a model system to show how this information is used to create the radiographs.
\Fig{test} shows a target partitioned into two regions of refractive index $n_1, n_2$ and opacities $\kappa_1, \kappa_2$.
When x rays are incident on the target normally, they travel along the direction from where they were launched until they are refracted because of a change in the refractive index.
Also, the rays are launched with an arbitrary intensity of 1; their final intensity of exp$\left[\dsp{-\int_0^L\kappa({\bf r}) ds}\right]$ is then determined based on their total path length $L$.
In this model system, randomly launched rays are incident normally on the rectangular slab shown in \fig{test}.
Three such representative ray trajectories through the three regions are shown. 
The rays in regions I and III do not get deflected, while the rays in region II get refracted by an angle of $\pi/3$ at the medium interface and get deflected from their path.
The angle of deflection $\theta$ plotted in the figure shows the degree of deflection from its launching point and not from the interface. 
Due to the geometry of the setup, $\theta$ is maximum near the interface of regions I and II where the rays have the largest path length after refraction.
This path length is decreased as we move from the interface of regions I and II to the interface of regions II and III and the angle $\theta$ decreases linearly.
\Fig{test} also shows the intensity representing the phenomena of refraction and attenuation.
When only refraction is considered, the model assigns a unit intensity to all the x-rays.
There are no rays accumulated near the region I and II interface as they deflected and get accumulated towards the end of region II, increasing the flux in that region.
If both refraction and attenuation are considered, the x-ray flux is weighted by the decrease in the intensity of the rays due to attenuation.
If we choose $\kappa_1=10$ cm$^{-1}$, the final intensity of the rays in regions I and III are $\exp(-0.2)$ and $\exp(-0.4)$ respectively. 
Through a derivation, the intensity in region II is found to be $\exp\left\{-0.1\left[2x-\frac{(2-x)}{\cos(\pi/12)}\right]\right\}$, where $x$ is the distance between the launch position of the ray and region I, II interface.
This result can also be derived by using Eq.~(10) of Ref.~\onlinecite{Koch_Fresnel_theory}.
Since the intensity is exponentially decreasing with $x$, this explains the increase in attenuation from the left to right in region II.
The analysis shown here for the model system was used to construct the simulated radiographs shown in this paper.

\bibliography{Xray_radiography_AKar} 

\end{document}